%% file: bcpnn_preprint.tex
\documentclass[runningheads]{llncs}
\usepackage[T1]{fontenc}
\usepackage{graphicx}
\usepackage{multirow}
\usepackage{wrapfig}
\usepackage[acronym, nopostdot]{glossaries}
\glsdisablehyper
\makeglossaries
\input{acronyms}
\usepackage{algorithm}
\usepackage{algpseudocode}
\usepackage{url}
\usepackage[dvipsnames]{xcolor}
\usepackage{breakurl}

\usepackage{graphicx}
\usepackage{array}
\usepackage{booktabs}
\begin{document}
\title{A Reconfigurable Stream-Based FPGA Accelerator for Bayesian Confidence Propagation Neural Networks}
\subtitle{Design, Implementation, and Performance Analysis}
\titlerunning{A Reconfigurable Stream-Based FPGA Accelerator for BCPNN}
%
\author{Muhammad Ihsan Al Hafiz\inst{1} \and
Naresh Ravichandran\inst{1} \and
Anders Lansner\inst{1,2,3} \and
Pawel Herman\inst{1,3,4,5} \and
Artur Podobas\inst{1,3}}
\authorrunning{M. I. A. Hafiz et al.}
%
\institute{KTH Royal Institute of Technology, Stockholm, Sweden \\ \and
Stockholm University, Stockholm, Sweden \\ \and
Swedish e-Science Research Centre (SeRC), Sweden \\ \and
Digital Futures, Stockholm, Sweden \\ \and
International Research Centre for Neurointelligence, University of Tokyo, Japan \\
\email{\{miahafiz, nbrav, ala, paherman, podobas\}@kth.se}}
\maketitle              
\begin{abstract}

Brain-inspired algorithms are attractive and emerging alternatives to classical deep learning methods for use in various machine learning applications. Brain-inspired systems can feature local learning rules, both unsupervised/semi-supervised learning and different types of plasticity (structural/synaptic), allowing them to potentially be faster and more energy-efficient than traditional machine learning alternatives. Among the more salient brain-inspired algorithms are Bayesian Confidence Propagation Neural Networks (BCPNNs).
BCPNN is an important tool for both machine learning and computational neuroscience research, and recent work shows that BCPNN can reach state-of-the-art performance in tasks such as learning and memory recall compared to other models. Unfortunately, BCPNN is primarily executed on slow general-purpose processors (CPUs) or power-hungry graphics processing units (GPUs), reducing the applicability of using BCPNN in (among others) Edge systems.
In this work, we design a custom stream-based accelerator for BCPNN using Field-Programmable Gate Arrays (FPGA) using Xilinx Vitis High-Level Synthesis (HLS) flow. Furthermore, we model our accelerator's performance using first principles, and we empirically show that our proposed accelerator is between 1.3x - 5.3x faster than an Nvidia A100 GPU while at the same time consuming between 2.62x - 3.19x less power and 5.8x - 16.5x less energy without any degradation in performance.

\end{abstract}
\section{Introduction}

\gls{DL}~\cite{lecun2015deep} architecture has emerged as one of the most essential machine learning tools in the past decades. \glspl{DL} are used in everything from image recognition~\cite{alzubaidi2021review} and time-series prediction~\cite{lindemann2021survey} to natural language processing~\cite{kalyan2021ammus}. Since their inception around 2012, the size of \gls{DL} systems has been growing at an exponential rate, demanding more and more computational resources and power~\cite{sevilla2022compute}. In particular the latter, energy consumption, has been identified as challenge to overcome since training a modern \gls{DL} system can take several months and can be very energy-consuming (ChatGPT4 consumed ~50 million kWh~\cite{jia2024analysis}). In short, there is a growing need to research alternative machine learning methods in order to satisfy performance demands without needlessly taxing the environment. One such direction is to draw inspiration from biology and investigate \texttt{brain-inspired} systems.

A \texttt{brain-inspired} system is a system that solves machine learning problems in a way abstracted but derived from theories of the brain in computational neuroscience. A brain-inspired system can be either spiking~\cite{maass1997networks} (often called neuromorphic system~\cite{schuman2017survey}) or rate-based (non-spiking). Brain-inspired systems typically have several traits that make them attractive to use: \textbf{(i)} they can be very sparse and energy-efficient, \textbf{(ii)} they have local (non-propagating) learning rules, \textbf{(iii)} supports one- and few-shot learning, and \textbf{(iv)} they can provide insight into how the brain computes. There are multiple brain-inspired machine learning models, but few are as salient and with such mature theory as the \textit{Bayesian Confidence Propagation Neural Network} (BCPNN)~\cite{bate1998bayesian}.

BCPNN is a biologically plausible model that is derived from the organization of the human cortex~\cite{mountcastle1997columnar}, where the basic building blocks are hypercolumns and minicolumns. BCPNN supports multiple different forms of learning, including learning of synaptic strengths~\cite{citri2008synaptic} (based on Bayes theorem) as well as structural plasticity~\cite{lamprecht2004structural} that rewire the connections between building blocks. More importantly, BCPNN supports supervised, semi-supervised, and unsupervised learning~\cite{ravichandran2021brain}, making it a strong choice for systems with a limited amount of labelled data. While BCPNN has shown state-of-the-art training and inference performance~\cite{ravichandran2024unsupervised} in multiple data sets using general-purpose \gls{CPU} and \gls{GPU} implementation, these devices are typically too expensive (e.g., in terms of power consumption) to deploy on Edge computing devices that could leverage the properties of BCPNN.

In this work, we propose the first high-performance hardware accelerator for BCPNN. We have described our data-flow accelerator using the Xilinx Vitis \gls{HLS}~\cite{nane2015survey} toolchain and executed it on state-of-the-art Alveo U55C Field-Programmable Gate Arrays (FPGAs). We claim the following contributions:

\vspace{-0.15cm}

\begin{enumerate}
\item We describe and implement the first high-performance BCPNN FPGA accelerator for use in data centers and edge systems that support both inference as well as online (unsupervised) learning in faster-than real-time,
\item we apply the BCPNN theory on two new data-sets: detecting Pneumonia and Breast cancer,
\item We develop an analytical performance model (based on first principles) to provide insight into the performance of our hardware accelerator and
\item We empirically quantify the performance of our accelerator, positioning it against an Tesla-class Nvidia A100 GPU and a Intel Xeon server-class CPU, showing an  advantage in both performance and power consumption of our FPGA accelerator
\end{enumerate}

\section{Related Work}

BCPNN has a long lineage of research work dating back to 1980s~\cite{lansner1989one}. Since then, several research works have extended the use of BCPNN to (among others) drug reaction signal generation~\cite{bate1998bayesian}, pattern recongition~\cite{orre2005bayesian}, spike-based formulation~\cite{tully2016spike}, investigated support for fixed-point arithmetic~\cite{johansson2004bcpnn}, and several machine learning applications~\cite{ravichandran2021brain,svedin2021higgs,ravichandran2022brain} and more. Motivated by the success and versatility of BCPNN, several groups have proposed hardware accelerators to improve performance and reduce the energy consumption of BCPNN. In 2020, Yang et al.~\cite{yang2020optimizing} optimized the BCPNN learning rule with respect to memory accesses, showing how non-coalesced column-wise memory access patterns in lazy-based methods can be eliminated, which can result in significant speed ups. In 2020, Liu et al. \cite{liu_fpga-based_2020} implemented an \gls{FPGA}-based hardware accelerator for a spiking-based \gls{BCPNN} model. This architecture employs a 'lazy update mode', efficiently updating eight local synaptic state variables by optimizing parallelism and decomposing calculations based on inherent data dependencies. These optimizations reduce the computation and bandwidth by more than two orders of magnitude, which makes efficient implementation of \gls{BCPNN} for real-time brain simulation engine \cite{liu_fpga-based_2020}. This approach led to a substantial acceleration in processing time, with an update time of 110 ns on an \gls{FPGA}, compared to 25800 ns on a \gls{CPU} \cite{liu_fpga-based_2020}. Podobas et al. introduced StreamBrain~\cite{podobas_streambrain_2021} in 2021, a framework that enables the deployment of the rate-based \gls{BCPNN} in High-Performance Computing (HPC) systems. StreamBrain is a domain-specific language (DSL) that supports various backends, including \glspl{CPU}, \glspl{GPU}, and \glspl{FPGA}. The authors demonstrate the practical capabilities of StreamBrain by training the MNIST dataset within seconds and to show the result of \gls{BCPNN} in higher-dimension problems with STL-10 networks. Additionally, the paper explores the use of custom floating-point formats and the impact when using \glspl{FPGA}. However, unlike the present paper, StreamBrain only accelerated a small subset of the BCPNN algorithm on the FPGA platform. Wang et al.~\cite{wang2021mapping} showed that the BCPNN local learning rule can be mapped and executed using an analog memristor model, showing that the device could have a correlation coefficient as high as $0.98$, and showing that it could learn the MNIST benchmark. Wang et al. \cite{wang_fpga-based_2024} presented an \gls{FPGA}-based HPC design specifically optimized for a \gls{BCPNN}-based associative memory system. Their approach incorporates several optimizations, including shared parallel computing units, hybrid-precision computing for a hybrid update mechanism, and the globally asynchronous, locally synchronous (GALS) strategy. Using the Xilinx Alveo U200 \gls{FPGA} accelerator card, the design achieved a maximum network size of 150x10 and a peak frequency of 100 MHz. The \gls{FPGA}-based solution outperformed its Nvidia GTX 4090 counterpart, demonstrating a maximum latency reduction of 33.23x and a power consumption reduction of over 6.9x. The study underscores the potential of \gls{FPGA}-based accelerators to significantly enhance both speed and energy efficiency in neuromorphic computing implementations. However, the scope of their work is limited to a small network size and omits evaluation of real-world datasets.
Contrary to the related work, which has been shown either in-parts~\cite{podobas_streambrain_2021,yang2020optimizing,wang2021mapping} or at a low TRL (omitting real use-cases)~\cite{wang_fpga-based_2024}, our work is the first that provides an FPGA accelerator for BCPNN that is high-performance (outperforms Nvidia A100) and that can handle real-life use-case with a low-latency, encourage its deployment in data-centers and on-edge premises. We are also the first to show that BCPNN with the (more complicated) use cases, such as detecting pneumonia or breast cancer.

\section{Bayesian Confidence Propagation Neural Network}

\gls{BCPNN} is a brain-inspired machine learning model that utilizes the principles of Bayesian statistics to derive the synaptic and neuronal update operations. It has two types of formulation: spike-based and rate-based. In this paper, we design a hardware accelerator for the rate-based \gls{BCPNN}. The work is based on the latest work in \cite{ravichandran2024unsupervised}, which is a feedforward \gls{BCPNN} that integrates cortical column, divisive normalization, Hebbian synaptic plasticity, structural plasticity, sparse activity, and sparse, patchy connectivity.  

The BCPNN divides its computational units into \textit{minicolumns}, which form part of larger modules known as \textit{hypercolumns} \cite{ravichandran2024unsupervised}. Each hypercolumn encodes a particular input attribute, while its constituent minicolumns represent discrete, mutually exclusive values of that attribute. This arrangement echoes the columnar structure of the primate neocortex, where functionally similar neurons are grouped vertically, creating a sparse and energy-efficient coding scheme \cite{douglas_canonical_1989,douglas_neuronal_2004}.

A basic feedforward BCPNN consists of at least two layers: an input layer and a hidden layer. The input layer’s minicolumns capture discrete feature values provided by the data, and the hidden layer’s minicolumns encode internal representations derived from these inputs \cite{ravichandran2024unsupervised}. Connecting these layers are weighted projections that undergo synaptic plasticity, an unsupervised learning mechanism analogous to Hebbian-Bayesian updates. This rule adapts the network parameters online, using local statistics of neuronal activities.

At the core of BCPNN lie three key probability traces, incrementally updated at each training step: the probability of an input minicolumn being active ($p_{i}$), the probability of a hidden minicolumn being active ($p_{j}$), and their joint probability ($p_{ij}$). These traces support a learning rule where biases ($b_{j}$) and connection weights ($w_{ij}$) are computed as logarithms of conditional probabilities:

\begin{equation} b_{j} = \log p_{j}, \quad w_{ij} = \log \frac{p_{ij}}{p_{i}p_{j}}. \end{equation}

This formulation expresses the hidden unit’s bias as the self-information and the synaptic weight as the mutual information between pre- and post-synaptic activities. Conceptually, it transforms observed co-occurrences of events into updated parameters that influence network activity. The activation of each hidden minicolumn is determined by a softmax function applied to support values derived from weighted input signals. This ensures that minicolumns in the same hypercolumn compete, resulting in a probability distribution across features. Consequently, a BCPNN hypercolumn provides a discrete probability estimate that closely resembles the cortical microcircuit behaviour, where excitatory and inhibitory interactions lead to sparse, distributed coding patterns. Finally (and importantly), BCPNN also supports structural plasticity where the network changes as a function of time, complementing the synaptic learning rule described above.

In short, BCPNN integrates neuroscientific principles—cortical microcircuitry, local learning, and probabilistic coding—into a neural computation framework. It encodes probability distributions directly within its weights and biases, learns online from streaming data, and yields a compact, high-level representation of complex inputs.

\section{High-Performance Stream-Based BCPNN Accelerator}

\begin{figure}
    \centering
    \includegraphics[width=9cm]{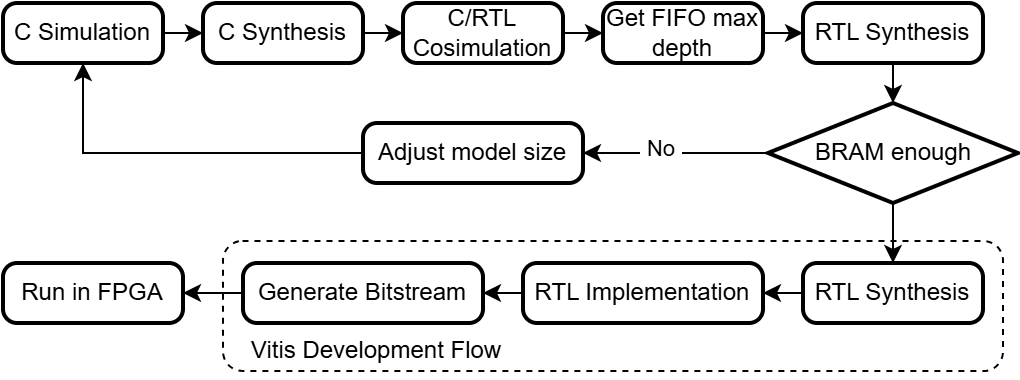}
    \caption{Design workflow}
    \label{fig:workflow}
\end{figure}

Figure \ref{fig:workflow} illustrates our complete development workflow. We start with a C-level simulation to verify correctness, then proceed to C-level synthesis to obtain a preliminary estimate of hardware resources. Next, we perform C/\gls{RTL} cosimulation to finalize \gls{FIFO} depths and confirm that no deadlocks can occur. If we encounter resource constraints, we adjust model sizes or parameters before moving on to \gls{RTL} synthesis for a more accurate assessment of hardware utilization. Once the design meets our resource and timing requirements, we transition to the Vitis development flow. This process packages the \gls{RTL} into an extensible platform, performs synthesis and implementation, and generates the \gls{FPGA} bitstream. By leveraging Vitis flow, we can concentrate on optimizing the \gls{BCPNN} kernel, as low-level tasks such as PCIe or DMA configuration are handled automatically.

\subsection{Accelerator Design using HLS}

The \gls{BCPNN} kernel comprises three interconnected population layers: input, hidden, and output. Each population layer represents a group of neurons that encodes and processes probabilistic relationships. These layers communicate through projection layers, with the input-hidden projection connecting the input population to the hidden population, and the hidden-output projection linking the hidden population to the output population. A projection refers to the connections in which information is transmitted from one population of neurons to another. To simplify \gls{FPGA} optimization, we set key dimensions (e.g., hidden layer sizes) to powers of two or multiples of four. This choice eases unrolling and data partitioning during \gls{HLS}.

\begin{figure}
    \centering
    \includegraphics[width=10cm]{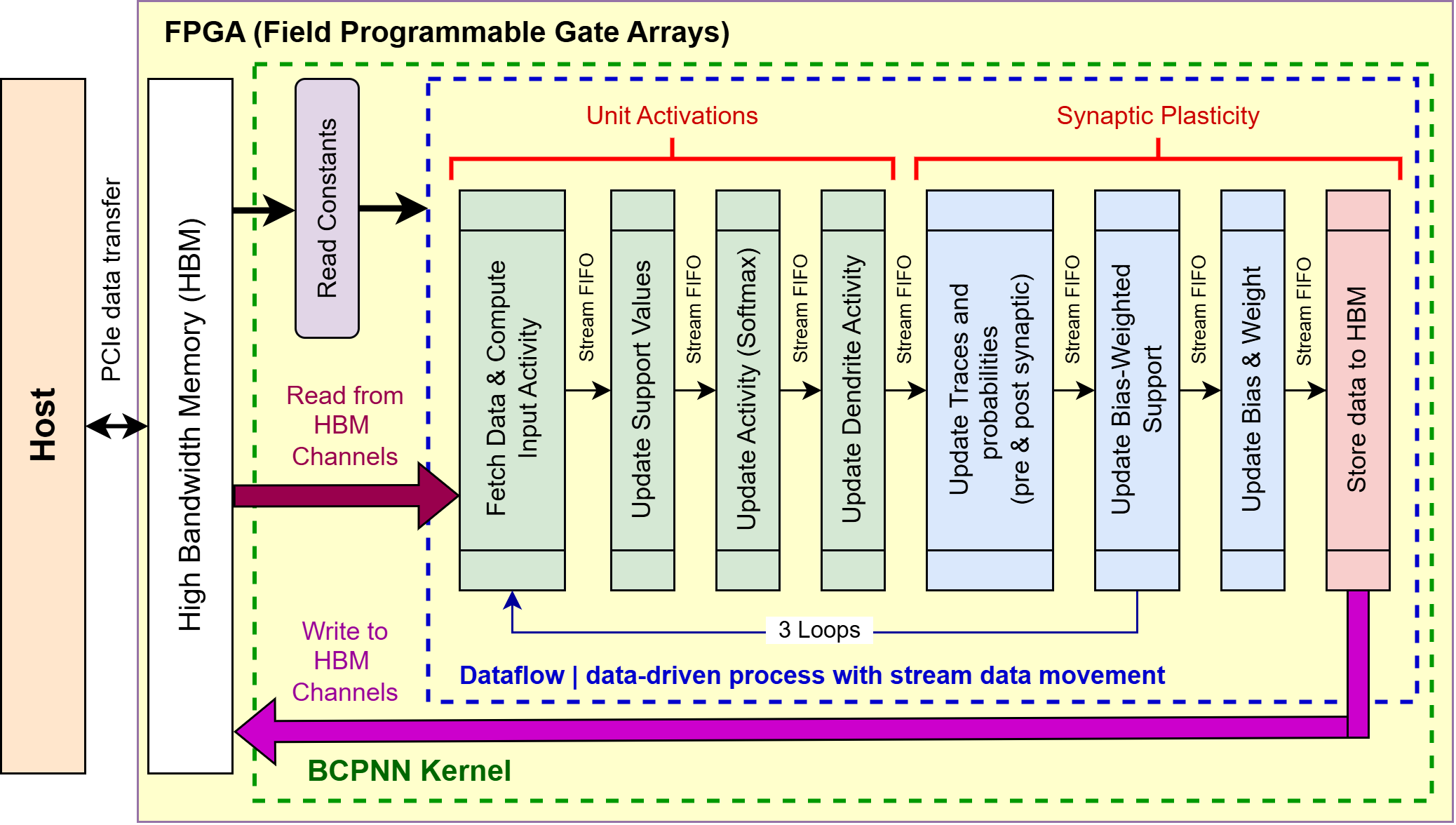}
    \caption{Block diagram connection host to FPGA}
    \label{fig:general}
\end{figure}

Building on these structural decisions, we designed the \gls{BCPNN} kernel as a stream-based, data-driven architecture as shown in Figure \ref{fig:general}. Starting from a C/C++ specification, the \gls{HLS} flow generates \gls{RTL} that covers both unit activations and synaptic plasticity, which are the most computationally demanding. Although some routines depend on each other’s outputs and thus execute sequentially, operations associated with separate populations and projections are inherently independent. This independence enables parallelization through multiple streaming pipelines.

Expanding on the range of functionalities, the complete kernel supports unsupervised, supervised, and inference modes, with or without structural plasticity. Although each mode reuses the same streaming pipeline and, therefore, might appear to have similar execution times, there is a key exception in the inference-only design. Inference does not require synaptic plasticity updates (weights, biases, and activity probabilities remain fixed), which reduces \gls{BRAM} usage and allows for higher clock frequencies. This inference-specific configuration is particularly beneficial for energy-sensitive edge deployments. Although the final kernel design takes advantage of parallel streaming and specialized inference-only configurations, this level of efficiency and resource utilization was not achieved in a single step. Our development process began with a more straightforward sequential implementation. Starting from this initial baseline allowed us to identify bottlenecks in computation and memory access, paving the way for the subsequent optimization strategies described below.

\subsubsection{Initial Unoptimized Sequential Implementation. }

\begin{figure}
    \centering
    \includegraphics[width=9cm]{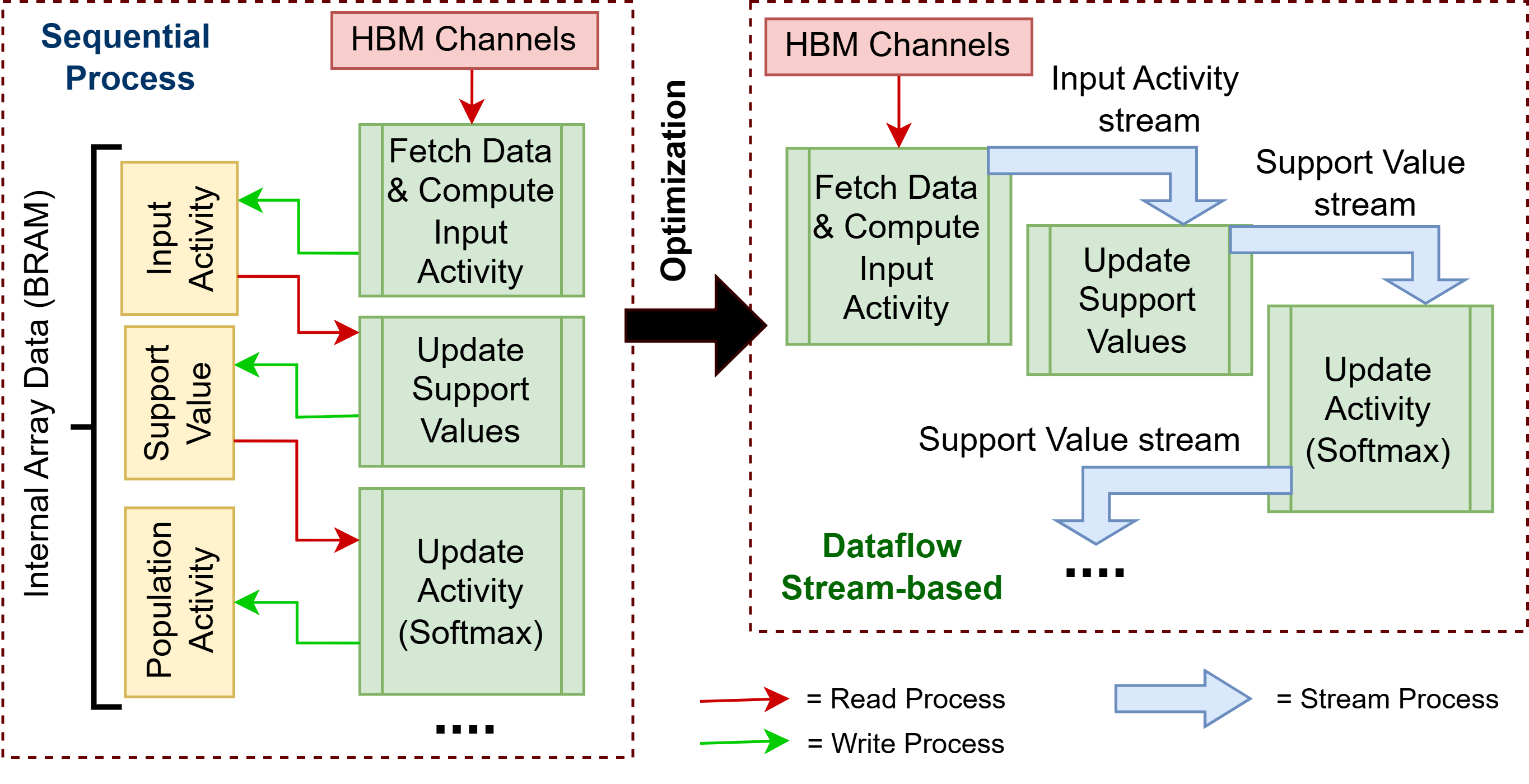}
    \caption{Optimization from sequential process to dataflow stream-based}
    \label{fig:dataflow}
\end{figure}

As illustrated in Figure \ref{fig:dataflow}, our initial implementation processed each subtask sequentially. This approach wasted resources because the hardware allocated for other steps remained idle during the execution of the current step. It also introduced challenges in handling data: storing all data on-chip consumed an excessive amount of \gls{BRAM} and led to routing congestion while relying on off-chip memory caused significant latency overhead. Recognizing these inefficiencies, we pursued several optimization techniques to enable parallelism, reduce memory overhead, and improve overall throughput.

\subsubsection{Optimization \#1: Stream-based FIFO data. }
The first step was to adopt a stream-based data transfer model, where data elements are packaged into fixed-size segments and forwarded continuously through \glspl{FIFO}. Rather than using static arrays in \gls{BRAM}, we defined \gls{FIFO} channels with a fixed depth to control data flow dynamically. We found that this approach reduces on-chip memory usage, mitigates routing complexity, and provides a foundation for task-level parallelism. However, streams alone are insufficient; we still need to break the sequential execution pattern.

\subsubsection{Optimization \#2: Dataflow process. } Dataflow directives in \gls{HLS} enable task-level pipelining, allowing multiple sub-tasks to run concurrently as long as they are not interdependent. As shown in Figure~\ref{fig:dataflow}, each stage of the computation can begin processing as soon as partial data is ready, passing intermediate results through \gls{FIFO} streams. This setup lets independent operations, such as those performed on different populations and projections, proceed in parallel, significantly increasing throughput. When combined with stream-based \glspl{FIFO}, dataflow introduces backpressure to maintain synchronization, preventing writes when \glspl{FIFO} are full and reads when they are empty. Certain operations, such as the softmax computation for updating activity levels, require waiting until all relevant data arrives. To avoid deadlocks and ensure every stage has the data it needs, we carefully size the \gls{FIFO} depths. Figure~\ref{fig:workflow} illustrates our systematic approach to determining optimal \gls{FIFO} configurations without resorting to trial and error. By applying dataflow directives alongside stream-based data transfers, our BCPNN kernel achieved roughly a 70\% performance improvement compared to the initial sequential implementation.

\subsubsection*{Optimization \#3: Spread memory mapping in HBM with data partitioning and data merging.}
\begin{figure}
    \begin{center}
    \includegraphics[width=9cm]{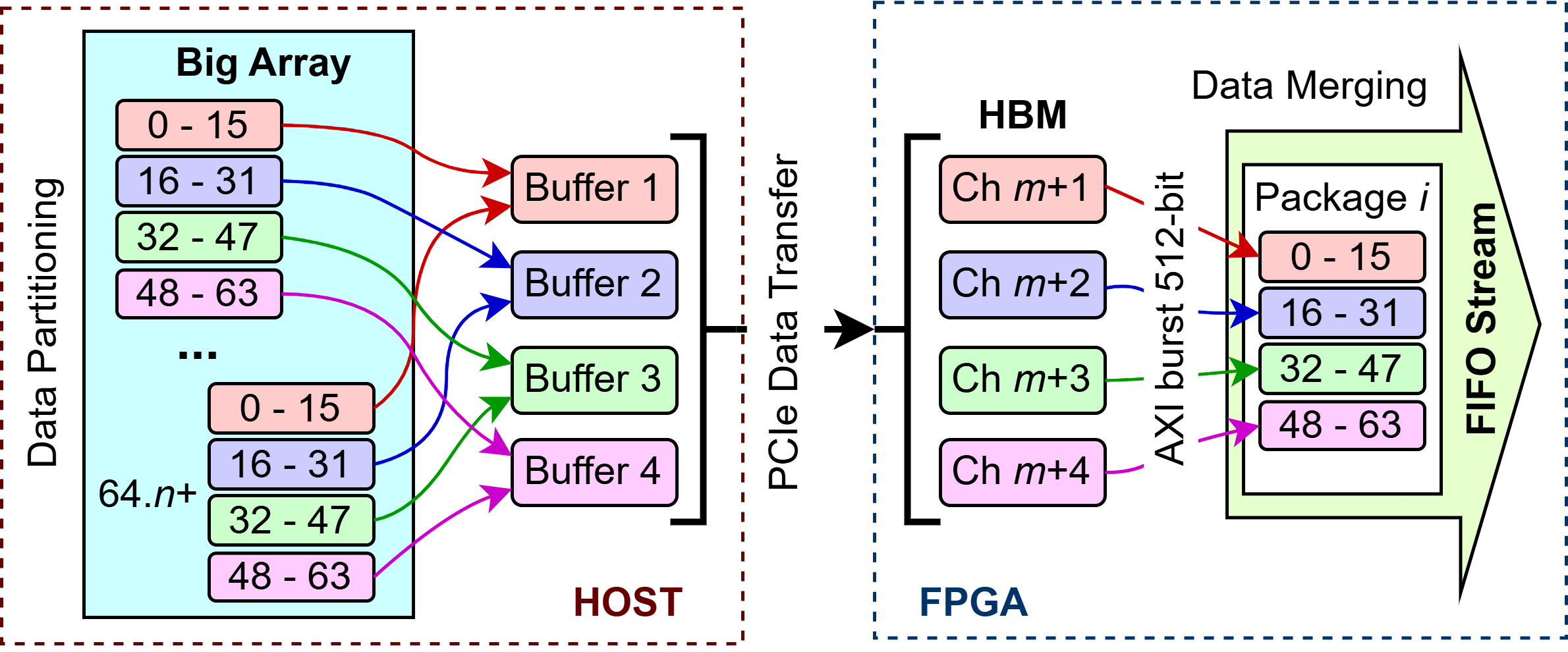}
    \end{center}
    \caption{Parallel HBM Access with Data Partitioning and Merging}
    \label{fig:hbmaccess}
\end{figure}

As shown in Figure~\ref{fig:hbmaccess}, we further improve performance by leveraging multiple HBM channels through data partitioning and merging. Large arrays from the input-hidden projection layer (e.g., joint probability and weight data) are divided into four segments, each streamed to a separate HBM channel. On the FPGA, we use 512-bit burst reads, equivalent to fetching 16 floating-point values at a time, from each channel. Although HBM natively supports 256-bit access, its higher frequency (450 MHz) allows this effective doubling to 512 bits at our lower clock rate (<300 MHz)~\cite{xilinx_inc_axi_2022}. The data from all four channels is then merged into a single stream packet of 64 floating-point values. Aligning indexing between pre-/post-synaptic activities allows these large packets to be processed in parallel using \gls{HLS} unroll directives. For the hidden-output projection, we apply a similar burst-read strategy without partitioning, producing 16-value packets to maintain efficient dataflow. Since the input-hidden and hidden-output projections operate in parallel, this optimization reduces latency by a factor of about 64. A similar approach is used for write operations.

\subsection{Performance Analysis}

We conducted an internal performance analysis to guide platform-specific optimizations. To accomplish this, we employ a roofline model that highlights bottlenecks and helps us refine the design for a given hardware.

The Roofline Model~\cite{williams_roofline_2009} helps us visualize whether an application is limited by compute resources or memory bandwidth. It does so by plotting achievable performance (in FLOP/s) against arithmetic intensity ($I$, defined as the ratio of floating-point operations to bytes of data moved). On conventional architectures, if $I$ is lower than the machine balance $M_b$, the application is memory-bound; otherwise, it is compute-bound~\cite{calore_fer_2022,john_mccalpin_memory_1995,siracusa_comprehensive_2022}.

Adapting this model to FPGAs is non-trivial. Unlike fixed architectures, an FPGA’s theoretical peak compute performance $C_{FPGA}$ depends on how many operations can be mapped onto its available resources and the operating frequency $f_{imp}$. We start with the number of available resources $R_A$ and the resource requirement per operation $R_O$. The ratio $R_A/R_O$ indicates how many such operations can run in parallel. Incorporating a utilization factor $U_{R}^i$ (to account for routing congestion and practical limits, often around 80\%), and the implemented frequency $f_{imp}$, we have:

\begin{equation} 
C_{FPGA} = f_{imp} \times \min_{i}\left(\frac{R_A^i}{R_O^i} \times U_{R}^i \right) 
\end{equation}

If we focus on DSPs and LUTs as the primary resources for floating-point operations, this simplifies to:

\begin{equation} 
C_{FPGA} = f_{imp} \times \min\left(\frac{R_A^{LUT}}{R_O^{LUT}} \times U_R^{LUT}, \frac{R_A^{DSP}}{R_O^{DSP}} \times U_R^{DSP}\right)
\end{equation}

Next, we determine the FPGA’s memory bandwidth $B_{HBM}$ by considering the HBM frequency $f_{HBM}$, data width $W_{HBM}$, and the number of channels $Ch_{HBM}$:

\begin{equation} 
B_{HBM} = f_{HBM} \times W_{HBM} \times Ch_{HBM}
\end{equation}

Finally, the machine balance $M_b$ for the FPGA is given by:

\begin{equation} 
    M_b = \frac{C_{FPGA}}{B_{HBM}} = \frac{f_{imp} \times \min\left(\frac{R_{A}^{LUT}}{R_{O}^{LUT}}  \times U_{R}^{LUT}, \frac{R_{A}^{DSP}}{R_{O}^{DSP}}  \times U_{R}^{DSP} \right)}{f_{HBM} \times W_{HBM} \times Ch_{HBM}} 
\end{equation}

By placing our kernel’s arithmetic intensity $I$ on the Roofline plot and comparing it to $M_b$, we can determine if it is operating in a memory-bound or compute-bound region for our particular FPGA implementation. This helps guide subsequent optimizations, either by increasing arithmetic intensity (e.g., reusing data to reduce memory traffic) or by improving the resource utilization and frequency (to push $C_{FPGA}$ closer to its theoretical peak).

\subsubsection{Theoretical Performance and Bandwidth.}

In this work, we implemented the kernel with single floating-point precision, albeit future work can easily use other number representations. The theoretical peak performance can be estimated by using a multiply-accumulation operation that consists of one addition and one multiplication. This method is similar to the evaluation in \cite{calore_fer_2022}. Based on the report resource utilization for floating-point by Xilinx \cite{xilinx_inc_performance_2023} for our \gls{FPGA}, the addition operation requires 192 LUTs and 2 DSPs, whereas the multiplication operation requires 74 LUTs and 3 DSPs. On the other hand, Xilinx Alveo U55C consists of 1146240 LUTs and 8376 DSPs. Therefore, the computation performance $C$ for frequency implementation 100 MHz with an assumption utilization maximum of 80\% is 288.77 GFLOPs/s. Moreover, the Xilinx Alveo U55C HBM has 32 pseudo channels with bit-width 256 and runs normally at 450 Mhz. so the maximum bandwidth of HBM is 460 GB/s \cite{xilinx_inc_axi_2022}.

\section{Experimental Setup}

We implemented the \gls{BCPNN} kernel with three distinct models, each with a different dataset and network configuration, to demonstrate its reconfigurability and adaptability (albeit, nothing limits our framework for creating accelerators with other models).  As shown in Table~\ref{tab:model_configs}, these models vary in terms of input dimension, hidden layer size, output classes, dataset scale, and the number of epochs used for unsupervised training. The parameter \textit{nactHi} defines the sparsity of the input for both with and without structural plasticity. In our approach, we adopt a semi-unsupervised setup: the epoch count listed pertains to the unsupervised training phase, while the supervised training phase is performed once per configuration. MNIST comprises 28x28 grayscale images of handwritten digits from 0 to 9 \cite{lecun_gradient-based_1998}. The Pneumonia and Breast dataset are the part of MedMNIST dataset \cite{yang_medmnist_2021,yang_medmnist_2023}. The Pneumonia dataset includes pediatric chest X-ray images and focuses on a binary classification task: distinguishing healthy (normal) cases from pneumonia-infected lungs \cite{yang_medmnist_2021}. The Breast dataset contains ultrasound images originally split into three classes (normal, benign, and malignant). For our binary classification, we combined normal and benign into a single positive category and treated malignant cases as negative \cite{yang_medmnist_2021}. \textit{This is the first time the BCPNN theory has been applied to the pneumonia and medical breast use-cases.}

\begin{table}[h]
 \vspace{-0.7cm}
    \centering
    \caption{Model Configurations and Dataset Details}
    \begin{tabular}{c|c|c|c|c|c|c|c|c|c}
        \textbf{Model} & \textbf{Dataset} & \textbf{Input size} & \multicolumn{2}{|c|}{\textbf{Hidden Layer}} & \textbf{nactHi} & \textbf{Out} & \multicolumn{2}{|c|}{\textbf{Data size}} & \textbf{Epoch} \\ \cline{4-5} \cline{8-9}
        & & & \textbf{Hyper} & \textbf{Mini} & & & \textbf{Train} & \textbf{Test} & \\ \hline \hline
        Model 1 & MNIST & 28x28 & 32 & 128 & 128 & 10 & 60000 & 10000 & 5 \\ \hline
        Model 2 & Pneumonia & 28x28 & 32 & 256 & 128 & 2 & 4708 & 624 & 20 \\ \hline
        Model 3 & Breast & 64x64 & 32 & 128 & 128 & 2 & 546 & 156 & 100 \\ \hline
    \end{tabular}
    \label{tab:model_configs}
    \vspace{-0.7cm}
\end{table}

To benchmark performance, we deployed our \gls{BCPNN} kernel on an AMD Xilinx Alveo U55C \gls{FPGA}, using the AMD Vitis Unified software platform v2023.2 and XRT v2.16.204. For comparison, we ran equivalent \gls{CPU} and \gls{GPU} implementations with identical configurations. The \gls{CPU} experiments were conducted on an Intel Xeon Silver 4514Y, compiled with g++ 11.4.0 and optimized using the -O3 flag on a single \gls{CPU} core. For the \gls{GPU}, we used Nvidia A100 and compiled it with CUDA 12.6.0 with optimization (-O3). We utilized the \gls{GPU} node from the High-Performance Computer Alvis\cite{noauthor_alvis_nodate}. Moreover, we compared the three implementations in terms of latency, power, and energy. \gls{CPU} power was not measured due to unsupported interfaces, and \gls{GPU} power was recorded using the visualization tools provided by the Alvis cluster \cite{noauthor_alvis_nodate}. The \gls{FPGA} measurements relied on real-time reporting through the XRT tool, ensuring accurate and direct observation of power usage.

Our reference implementation, written by domain experts, is in C/C++ with a CUDA backend for GPU acceleration. We modified the code to rely solely on the BCP layer for supervised learning and selected model sizes that align both with \gls{FPGA} resource constraints and dataset requirements. The \gls{GPU} implementation was similarly optimized using standard techniques and restricted to a single \gls{GPU}, ensuring a balanced comparison and a clear understanding of the efficiency gains offered by our \gls{FPGA}-based solution. 

We employed the implementation strategy \textit{Performance\_BalanceSLRs} to distribute logic evenly across the three \gls{SLR} of the \gls{FPGA}, mitigating routing congestion and improving achievable clock frequencies. The \gls{FPGA} that we used has three \gls{SLR}. Moreover, frequency selection was an iterative process, where resource utilization and routing complexity influenced the final operating speed. 

\section{Result}

Next, we evaluate our contributions in four areas: correctness, performance, analysis, and resource utilization.

\subsection{Correctness}

\begin{figure}[hbt!]
    \centering
    \includegraphics[trim={0 0 0 1cm},clip,width=12cm]{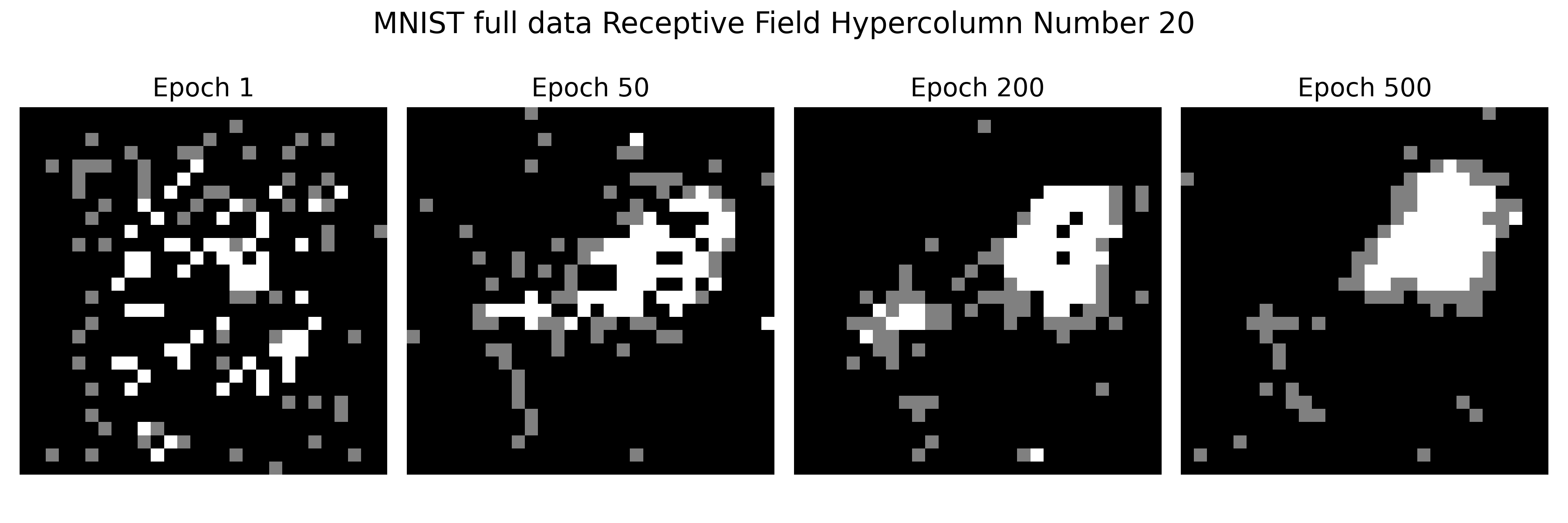}
    \caption{Structural plasticity can modify a hypercolumns receptive (or visual) field to extract most information from the data. Here we show how one receptive field in a HC change as a function of time, from a random (left) to a more refined (right) field.}
    \label{fig:struct}
\end{figure}

As shown in Table~\ref{tab:model_comparison}, the FPGA implementation achieves virtually the same accuracy as the reference CPU and GPU versions, confirming that the stream-driven dataflow architecture preserves the correctness of the C++ reference code. Across all models, accuracy differences are negligible, typically fractions of a percentage point. These minor discrepancies are primarily due to compiler optimizations (e.g. \texttt{unsafe-math-optimizations}) and slight variations in random number generation. Such factors can introduce small floating-point rounding differences and nonidentical data sampling patterns compared to CPU and GPU platforms.
Importantly, these variations do not affect the underlying BCPNN algorithm or its probabilistic learning rules. The FPGA-based accelerator still accurately replicates the intended model behaviour. Moreover, the important takeover, the test accuracy for the Pneumonia and Breast dataset is comparable with the accuracy from the CNN-based models that are reported in \cite{yang_medmnist_2021}. Figure~\ref{fig:struct} shows the receptive field of one HC and how it evolves with time, indicating that structural plasticity works as intended and in line with prior work~\cite{ravichandran_brain-like_2021}.

\newcommand{\rot}[2]{\multirow{#1}{*}{\rotatebox{90}{#2}}}
\definecolor{mygreen}{rgb}{0.0, 0.5, 0.0}
\definecolor{myred}{rgb}{0.5, 0.0, 0.0}
\newcommand{\blue}[1]{\textcolor{mygreen}{#1}}
\newcommand{\red}[1]{\textcolor{myred}{#1}}

\subsection{Performance}
\begin{table}[h!]
 \centering
\caption{Comparison of Model Implementations on Different Platforms (\textbf{infer}=inference only, \textbf{train}=w/training, \textbf{struct}=w/train+structural plasticity,\textbf{acc.}=accuracy, \textbf{-}=not available)}
\begin{tabular}{c|c l c c c c |c}

\textbf{Model} & \textbf{Type} & \textbf{Metric} & \textbf{Unit} & \textbf{CPU} & \textbf{GPU} & \textbf{FPGA} & \textbf{Impr.(over GPU)} \\
\hline \hline
\rot{10}{Model 1}& \rot{2}{Infer} & Latency    & ms & 2.644  & 1.495 & 0.280 & \blue{+5.3x}\\ 
                 &                & Energy/img & mJ & -      & 124.4 & 7.5   & \blue{+16.5x}\\ \cline{2-8}
                 & \rot{3}{Train} & Latency    & ms & 13.610 & 1.497 & 0.422 & \blue{+3.54x}\\
                 &                & Energy/img & mJ & -      & 124.6 & 11.3  & \blue{+11.02x}\\
                 &                & Total time & s  & 4302.9 & 572.2 & 314.9 & \blue{+1.81x} \\ \cline{2-8}
                 & \rot{3}{Struct}& Latency    & ms & 40.362 & 1.520 & 0.508 & \blue{+2.99x}\\ 
                 &                & Energy/img & mJ & -      & 126.5 & 13.7  & \blue{+9.23x}\\
                 &                & Total time & s  & 13286.8 & 621.6 & 473.9 & \blue{+1.31x}\\ \cline{2-8}  
                 & \rot{3}{Other} & Train acc. & \% & 94.5   & 94.6  & 94.5  & - \\
                 &                & Test acc.  & \% & 94.6   & 94.5  & 94.5  & - \\ 
                 &                & Power (W)  &    & -      & 83.2  & 27.0  & \blue{-3.08x} \\
\hline \hline
\rot{10}{Model 2}& \rot{2}{Infer} & Latency    & ms & 4.721  & 1.633  & 0.504 & \blue{+3.24x}\\
                 &                & Energy/img & mJ & -      & 146.6  & 14.2  & \blue{+10.32x}\\ \cline{2-8}
                 & \rot{3}{Train} & Latency    & ms & 27.4   & 1.646  & 0.552 & \blue{+3.03x}\\
                 &                & Energy/img & mJ & -      & 147.8  & 15.5  & \blue{+9.53x}\\ 
                 &                & Total time & s  & 2608.5 & 166.1  & 126.7 & \blue{+1.31x}\\ \cline{2-8}
                 & \rot{3}{Struct}& Latency    & ms & 55.258 & 1.631  & 0.609 & \blue{+2.63x}\\ 
                 &                & Energy/img & mJ &      - & 146.5  & 17.1  & \blue{+8.56x}\\
                 &                & Total time & s  & 5333.3 & 174.9  & 234.3 & \red{-0.75x}\\ \cline{2-8}  
                 & \rot{3}{Other} & Train acc. & \% & 91.5   & 91.0   & 91.5  & - \\
                 &                & Test acc.  & \% & 85.4   & 85.6   & 85.3  & - \\ 
                 &                & Power (W)  &    & -      & 89.8   & 28.1  & \blue{-3.19x}\\
 \hline \hline
\rot{10}{Model 3}& \rot{2}{Infer} & Latency    & ms & 2.649  & 1.541  & 0.540 & \blue{+2.75x}\\
                 &                & Energy/img & mJ & -      & 105.4  & 14.1  & \blue{+7.48x}\\ \cline{2-8}
                 & \rot{3}{Train} & Latency    & ms & 13.507 & 1.554  & 0.702 & \blue{+2.11x}\\
                 &                & Energy/img & mJ & -      & 106.3  & 18.3  & \blue{+5.8x}\\
                 &                & Total time & s  & 740.4  & 87.3   & 66.9  & \blue{+1.30x} \\ \cline{2-8}
                 & \rot{3}{Struct}& Latency    & ms & 38.319 & 1.556  & 0.690 & \blue{+2.26x}\\ 
                 &                & Energy/img & mJ & -      & 106.4  & 18.0  & \blue{+5.91x}\\
                 &                & Total time & s  & 2107.6 & 91.6   & 95.1  & \red{-0.96x}\\ \cline{2-8}  
                 & \rot{3}{Other} & Train acc. & \% & 89.1   & 89.7   & 89.7  & - \\
                 &                & Test acc.  & \% & 76.9   & 80.1   & 80.1  & - \\ 
                 &                & Power (W)  &    & -      & 68.4   & 26.1  & \blue{-2.62x} \\

\end{tabular}
\label{tab:model_comparison}
\vspace{-10pt}
\end{table}

Table \ref{tab:model_comparison} compares the performance of each model across \gls{CPU}, \gls{GPU}, and \gls{FPGA} platforms. The primary focus is on execution time, energy and power consumption. The \gls{FPGA} implementation consistently achieves a lower average processing time per image (latency) for both training and inference compared to the \gls{CPU} and \gls{GPU} in all the models, reducing total execution time, or the time for executing unsupervised training with the defined epoch, one supervised training, and inference for training and testing data. Total time execution has lower improvement than the latency because it has overhead from data transfer from host to FPGA and vice versa. When the model runs with structural plasticity, every certain training computes the structural plasticity that happens in the host, which significantly adds more overhead time. It affects more when it has a small dataset; then the structural plasticity process will be relatively more frequent than the bigger dataset. That is the reason why models 2 and 3 have a slightly slower total time for the structural plasticity version compared to the GPU. However, it does not affect bigger datasets like in model 1, when the total time for structural plasticity still outperforms GPU. 

In terms of power and energy consumption, the \gls{FPGA} demonstrates a substantial advantage over the \gls{GPU}. Whereas the \gls{GPU} draws between 68.4-89.8 W, the \gls{FPGA}’s power usage hovers around 26.1–28.1 W. Energy consumptions are reduced even more for all models and versions from 5.8x to 16.5x improvement over the GPU. This efficiency, combined with competitive performance, underscores the \gls{FPGA}’s suitability for energy-constrained environments. With significantly lower power consumption, the implementation of \gls{BCPNN} in an \gls{FPGA} with stream-based reconfigurable architecture indicates the promising possibility of applying it to edge applications. 

\subsection{Analysis}
\begin{figure}[hbt!]
    \centering
    \includegraphics[width=10cm]{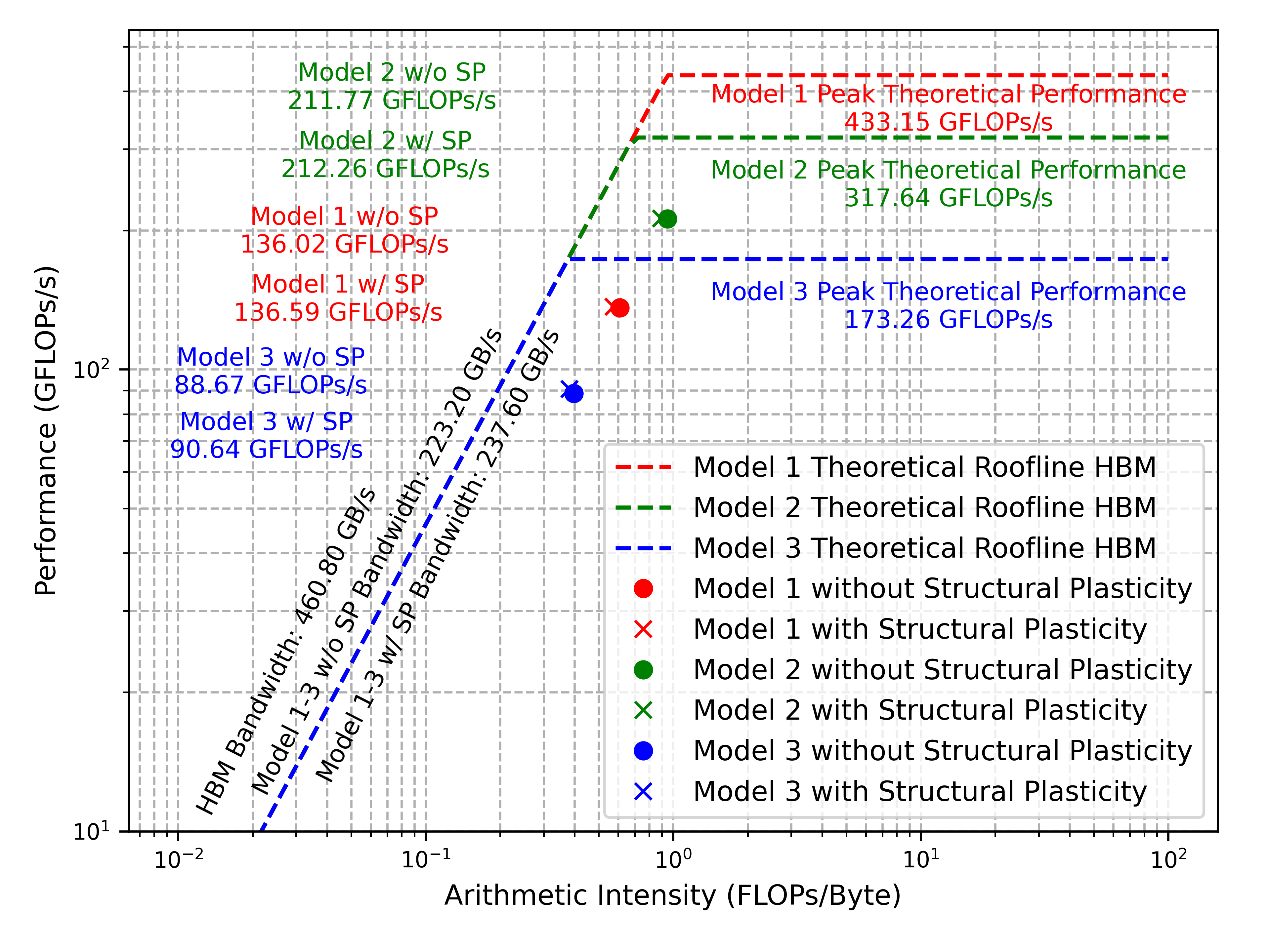}
    \caption{Roofline model plot of our accelerators (for the different models), showing performance (y-axis) as a function of arithmetic intensity (x-axis) for our accelerators, revealing how optimized our accelerators are (given theoretical upper limit).}
    \label{fig:roofline}
\end{figure}

Figure \ref{fig:roofline} illustrates the roofline model for the three \gls{BCPNN} models that are implemented with stream-based \gls{FPGA} implementation, both with and without structural plasticity. It provides valuable insights into the computational performance and memory access efficiency of the three \gls{FPGA}-based models. Every model's peak performance is derived with the assumption of maximum 80\% utilization for LUT and DSP with its operating frequency. It shows only for the full version of \gls{BCPNN} model implementation. None of the models achieve peak performance due to less than 80\% resource usage and specific algorithmic constraints. The design process has optimized the flow of data with stream-based \gls{FIFO} to make sure every resource will be maximally occupied during the operation. Using data partitioning for big arrays that are mapped to 4 pseudo channel HBM, we have \gls{BCPNN} to push the models up to the peak performance. However, since there is a necessity for operation in the \gls{BCPNN} algorithm to accumulate some arrays in some operations, the performance is limited. 

Model 1 has less actual performance than peak performance compared to the other models; it is because model 1 utilizes fewer hardware resources than the others. Overall, the reconfigurable design has been optimized using data-driven and array partitioning techniques in HBM. We limit the partition array in HBM to four because if we partition more, it will result in highly congested routing. 

Model 2 and Model 3 have better actual performance because they utilize more hardware resources. Model 2 lies closer to peak performance because the model size combination allows it to have very high floating point operation, while the stream-based method keeps the hardware utilization optimal. On the other hand, model 3 has lower peak performance because it can only be compiled with 60 MHz because the big input image requires a large allocation on the \gls{FIFO}, which results in high \gls{BRAM} utilization. 

The model with structural plasticity has a slightly bigger computation performance because it has additional computation for a sparsity array from the receptive field. It adds the need for bandwidth with an additional channel in HBM for 14.4 GB/s compared to the model without structural plasticity. 

In summary, the \gls{FPGA}-based \gls{BCPNN} implementation balances resource constraints and computational efficiency through a dataflow streaming approach and memory partitioning strategies. While the \gls{FPGA} may not always outperform a well-optimized \gls{GPU} in total execution time for every model, it consistently delivers lower power consumption and often competitive or superior per-image processing rates. The roofline analysis confirms that while current optimizations have moved the design closer to its theoretical limits, some algorithmic and architectural constraints remain. Addressing these constraints, such as exploring different partitioning factors or optimizing data access patterns, may further improve performance and resource efficiency in future implementations.

\subsection{Resource Consumption}
We evaluated the three versions of resource consumption of the \gls{BCPNN} kernel from every model. The first version is a full-featured kernel supporting unsupervised, supervised, and inference modes but without structural plasticity. The second version is a full-featured kernel with structural plasticity. The third version is an inference-only kernel. The inference kernel’s reduced complexity enables higher operating frequencies and lower resource utilization. This design choice makes it suitable for edge applications, where hardware resources, power, and execution time are often constrained.

\begin{table}[h]
\caption{FPGA Utilization (\textbf{infer}=inference only, \textbf{train}=w/training, \textbf{struct}=w/train+structural plasticity)
}
    \centering
    \renewcommand{\arraystretch}{1.2} 
    \setlength{\tabcolsep}{3pt} 
    \begin{tabular}{l|c|c|c|c|c|c}
        \textbf{} & \textbf{Version} & \textbf{LUT} & \textbf{FF} & \textbf{DSP} & \textbf{BRAM} & \textbf{Frequency} \\
        \hline  \hline
        \multirow{3}{*}{\rotatebox{90}{Model 1}} 
        & Infer & 174400 (15\%) & 257462 (11\%) & 550 (7\%) & 327.5 (18\%) & 200.0 MHz \\
        & Train & 454024 (40\%) & 546419 (24\%) & 3573 (43\%) & 437.5 (25\%) & 150.0 MHz \\
        & Struct & 475074 (41\%) & 574657 (25\%) & 3765 (45\%) & 473.5 (27\%) & 147.3 MHz \\
        \hline
        \multirow{3}{*}{\rotatebox{90}{Model 2}} 
        & Infer & 177201 (15\%) & 261754 (11\%) & 644 (8\%) & 701.5 (40\%) & 160 MHz \\ 
        & Train & 459419 (40\%) & 488973 (21\%) & 3573 (43\%) & 862.5 (49\%) & 110 MHz \\
        & Struct & 479801 (42\%) & 513057 (22\%) & 3765 (45\%) & 898.5 (51\%) & 107.8 MHz \\
        \hline
        \multirow{3}{*}{\rotatebox{90}{Model 3}} 
        & Infer & 180365 (16\%) & 259592 (11\%) & 640 (8\%) & 1419 (80\%) & 84.4 MHz \\
        & Train & 463580 (40\%) & 406798 (18\%) &  3573 (43\%) & 1568.5 (88\%) & 60.0 MHz \\
        & Struct & 481731 (42\%) & 430927 (19\%) & 3765 (45\%) & 1604.5 (90\%) & 60.0 MHz \\
    \end{tabular}
    \label{tab:util}
     \vspace{-0.7cm}
\end{table}

Table \ref{tab:util} presents the \gls{FPGA} utilization for the three models evaluated, offering a clear comparison of their resource demands. Among the three, the inference-only kernel stands out, consuming fewer resources and achieving higher operating frequencies compared to the full kernel. This highlights its effectiveness and suitability for edge application scenarios, where inference speed and hardware efficiency are critical. Notably, the addition of the structural plasticity feature introduces a slight increase in resource consumption, demonstrating the trade-off between utilizing the feature and resource efficiency.

The resource utilization scales with model complexity. For example, Model 2’s larger minicolumn in the hidden layer necessitates a fair increase in LUTs, FFs, and DSPs, and a more pronounced rise in \gls{BRAM} usage. Comparing Model 1 and 3, we see that increasing the input size from 28x28 to 64x64 significantly raises \gls{BRAM} utilization due to the need to buffer and process larger input data streams. Even though the architecture uses a stream-based design, certain operations require preloading data, resulting in higher on-chip memory usage.

\section{Conclusion}
In this paper, we introduced a reconfigurable stream-based FPGA accelerator for Bayesian Confidence Propagation Neural Networks (BCPNN), demonstrating its viability as a high-performance, energy-efficient platform for neuromorphic computing. \textit{This accelerator is currently the most power-efficient and fastest single-node implementation of the BCPNN theory}, opening up opportunities in deploying BCPNN for use in edge computing use-case as well as exploring computational neuroscience aspects of the theory. We achieved substantial gains over CPU and GPU equivalent implementations by leveraging a range of optimizations, such as stream-based \gls{FIFO}, dataflow parallelization, and strategic HBM channel partitioning. 
We evaluated our accelerator using three BCPNN model sizes across MNIST, Pneumonia, and Breast Medical datasets. In all cases, the FPGA-based system maintained comparable accuracy while substantially reducing latency, power, and energy usage. For example, on the MNIST dataset, the training time per image decreased from 1.497 ms on the GPU to 0.422 ms on the FPGA, and energy consumption for the train fell from 124.6 mJ to 11.3 mJ. Similar improvements were observed for the other datasets, underscoring the robustness and scalability of our design.
Overall, our FPGA accelerator achieves speedups of 1.3x to 5.3x compared to an NVIDIA A100 GPU, while simultaneously cutting power consumption by 2.62x to 3.19x and energy usage by 5.8x to 16.5x. These results indicate that an optimized FPGA-based approach can extend BCPNN deployments into resource-constrained environments where power, energy, and latency are critical. By bridging neuromorphic principles with specialized hardware design, this work moves brain-inspired models closer to real-world applications in edge and energy-sensitive systems.

\begin{credits}
\subsubsection{\ackname} This work was funded by the European Commission Directorate-General for Communications Networks, Content and Technology grant no. 101135809 (EXTRA-BRAIN), the Swedish Research Council grant no. 2021-04579 (Building Digital Brains), and the Swedish e-Science Research Centre (SeRC). The computations were enabled by resources provided by the Chalmers e-Commons at Chalmers and National Academic Infrastructure for Supercomputing in Sweden (NAISS), partially funded by the Swedish Research Council through grant agreement no. 2022-06725.

\subsubsection{\discintname} The authors have no competing interests to declare that are relevant to the content of this article.
\end{credits}
%
%
%
 \bibliographystyle{splncs04}
 \bibliography{final}

%
%
%
%
%
\end{document}

%% file: acronyms.tex
\newacronym{FMA}{FMA}{Fused Multiply Accumulate}
\newacronym{SIMD}{SIMD}{Single Instruction Multiple Data}
\newacronym{FPGA}{FPGA}{Field Programmable Gate Array}
\newacronym{ASIC}{ASIC}{Application Specific Integrated Circuit}
\newacronym{CPU}{CPU}{Central Processing Unit}
\newacronym{GPU}{GPU}{Graphics Processing Unit}
\newacronym{HLS}{HLS}{High Level Synthesis}
\newacronym{RTL}{RTL}{Register Transfer Level}
\newacronym{VHDL}{VHDL}{VHSIC Hardware Description Language}
\newacronym{BCPNN}{BCPNN}{Bayesian Confidence Propagation Neural Network}
\newacronym{BCP}{BCP}{Bayesian Confidence Propagation}
\newacronym{FIFO}{FIFO}{First In First Out}
\newacronym{HBM}{HBM}{High Bandwidth Memory}
\newacronym{PCIe}{PCIe}{Peripheral Component Interconnect Express}
\newacronym{AXI}{AXI}{Advanced eXtensible Interface}
\newacronym{BRAM}{BRAM}{Block Random Access Memory}
\newacronym{DMA}{DMA}{Direct Memory Access}
\newacronym{XRT}{XRT}{Xilinx Runtime library}
\newacronym{SLR}{SLR}{Super Logic Region}
\newacronym{NAISS}{NAISS}{National Academic Infrastructure for Supercomputing in Sweden}
\newacronym{DL}{DL}{Deep Learning}
\newacronym{HCU}{HCU}{hypercolumn unit}
\newacronym{DSP}{DSP}{Digital Signal Processor}
\newacronym{LUT}{LUT}{Look-Up Table}
\newacronym{TRL}{TRL}{Technology Readiness Level}